\documentclass[journal=jctcce,manuscript=article,layout=traditional]{achemso} % {traditional, twocolumn}
\usepackage[T1]{fontenc}
\usepackage[version=3]{mhchem}
\usepackage[dvipsnames]{xcolor}
\usepackage{graphicx,subcaption}
\usepackage{amssymb}
\usepackage{overpic}
\usepackage{braket}
\usepackage{soul}
\usepackage{bm}
\usepackage{hyperref}
\usepackage{enumerate}
\hypersetup{pdfborderstyle={/S/U/W 0.5}}
\newcommand{\pp}[2]{\frac{\partial{#1}}{\partial{#2}}}
\newcommand{\grad}[1]{\nabla_{\bm{#1}}}
\author{Tian Qiu}
\email{tianq@princeton.edu}
\affiliation{Department of Chemistry, Princeton University, Princeton, NJ 08540, USA.}
\author{Joseph E. Subotnik}
\affiliation{Department of Chemistry, Princeton University, Princeton, NJ 08540, USA.}
\title[]{An Efficient Algorithm for Constrained CASSCF(1,2) and CASSCF(3,2) Simulations as Relevant to Electron and Hole Transfer Problems}
\begin{document}
\maketitle

\begin{abstract}
We propose an efficient algorithm for the recently published electron/hole-transfer Dynamical-weighted State-averaged Constrained CASSCF (eDSC/hDSC) method studying charge transfer states and D$_1$-D$_0$ crossings for systems with odd numbers of electrons. By separating the constrained minimization problem into an unconstrained self-consistent-field (SCF) problem and a constrained non-self-consistent-field (nSCF) problem, and accelerating the direct inversion in the iterative subspace (DIIS) technique to solve the SCF problem, the overall computational cost is reduced by a factor of 8 to 20 compared with directly using sequential quadratic programming (SQP). This approach should be applicable for other constrained minimization problems and, in the immediate future, once gradients are available, the present eDSC/hDSC algorithm should allow for speedy non-adiabatic dynamics simulations.
\end{abstract}

\section{Introduction}
Charge transfer (CT) processes play a fundamental role in a wide range of chemical phenomena such as electrochemical reactions, redox reactions, and photo-chemical reactions. It is also a key factor in the design and development of novel catalysts, sensors, and energy storage systems.\cite{marcus:1956,Hush:1961,Levich:1961,Cukier:1998,Jortner:1998,Dennerl:2010} The theoretical investigation of CT is therefore of significant importance for understanding, predicting, and ultimately controlling the behavior of these diverse chemical systems\cite{cave:1996:gmh,Adams2003,blumberger:2017:transport:cr}.
The study of CT is challenging, however,  due to the need to simultaneously treat both electronic and nuclear degrees of freedom spanning a wide range of time scales. In particular, the coupled motion of nuclei and electrons is often  non-adiabatic and cannot be adequately described by the Born-Oppenheimer (BO) approximation\cite{martinez:1996:jpc}, instead requiring dynamical techniques to move along complicated potential energy surfaces (PES) with multiple avoided crossings and conical intersections\cite{hynes:ci_fssh:review}. 

In practice, the biggest challenge today in describing CT lies in the proper treatment of electron correlation. On the one hand, single-reference methods such as Hartree-Fock (HF) or Density Functional Theory (DFT) are inadequate for this purpose, as they fail to capture the multi-reference character of the wavefunction.\cite{martinez:2006:ci_topology_wrong} On the other hand, multi-reference methods such as Complete Active Space Self-Consistent Field (CASSCF)\cite{roos:1980,siegbahn:1981} and Multi-Reference Configuration Interaction (MRCI)\cite{grimme:1999,martinez:1996:jpc} can provide a more accurate description of CT.  Recent developments in CASSCF have made enormous progress as far as attacking larger and large systems\cite{sandeep:2017:casscf}, including through the use of density matrix renormalization group (DMRG) theory, and derivative couplings are available today for some multi-state CAS calculations\cite{shiozaki:2015:gradient} -- but the computational cost remains steep, which inevitably limits the usage of these methods if we seek to run dynamics on large systems. In short, one needs to find a method that balance the accuracy and efficiency for calculating electronic states and potential energy surfaces (PES) for CT. While constrained DFT (cDFT)\cite{wu:2006:cdft,wu:2007:cdftci,haobin:2010:cdft}  might be the standard choice today for such problems, derivative couplings do not yet exist and the method works best for weakly coupled systems\cite{troy:2015:cdftfail} (whereas we would like to work in both the strongly and weakly coupled limits). Please see below for a detailed discussion.

\begin{figure*}[ht!]
    \centering
    \includegraphics[width=0.85\textwidth]{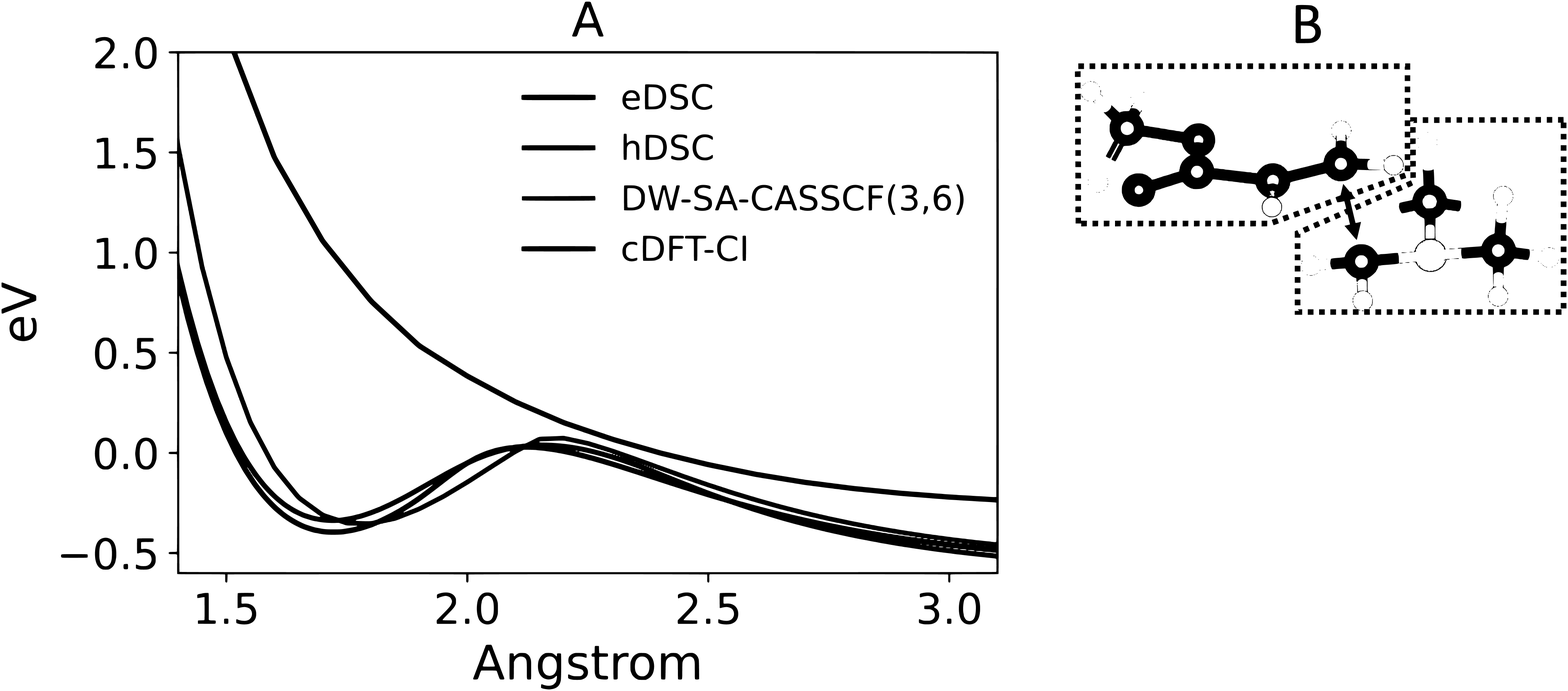}
    \caption{Ground PES for the bond dissociation of aminoalkyl radical-methyl acrylate system as predicted by eDSC, hDSC, cDFT-CI, and dynamically-weighted state-averaged CASSCF(3,6) (DW-SA-CASSCF(3,6)). (A) Ground PES predicted by different methods. Data in this figure are generated using methods from Ref. \citenum{Qiu:2024:dsc}. (B) Geometry of the aminoalkyl radical-methyl acrylate system. The black arrow indicates the horizontal axis of Fig. \ref{fig:dsc_vs_cdft}A.}
    \label{fig:dsc_vs_cdft}
\end{figure*}

In a recent paper, we have proposed an electron/hole-transfer Dynamical-weighted State-averaged Constrained CASSCF method (which we refer to as {\bf eDSC/hDSC}) \cite{Qiu:2024:dsc} tailored for studying CT processes in systems with an odd number of electrons.  Rather than work with the corresponding closed shell anion or cation as a reference\cite{weitao:2019:excited:dft}, we work directly with the radical system.  By imposing the constraints that the vector space spanned by active orbitals projects equally onto two pre-defined molecular fragments (i.e., donor and acceptor), the PES generated by our method is smooth and comparable to a much more computationally demanding CASSCF(3,6) calculation, for both weakly coupled and strongly coupled systems. As an example of the power of the method, in Fig. \ref{fig:dsc_vs_cdft}A we plotted the ground PES predicted by eDSC/hDSC and cDFT-CI for the bond dissociation of aminoalkyl radical-methyl acrylate system, which is reported to be a strongly coupled CT process (the geometry of this system is illustrated in Fig. \ref{fig:dsc_vs_cdft}B where the black arrow indicates the reaction path). The results show that, for systems with strongly coupled CT states, cDFT-CI fails to predict the bound ground state energy landscape but eDSC/hDSC can correctly predict the ground adiabat surface and the strongly coupled CT process. More discussions on this comparison can be found in Ref.
 \citenum{Qiu:2024:dsc}.

Now, while the cost of eDSC/hDSC is significantly lower than that of CASSCF(3,6) and one finds fewer minima (so the potential energy surfaces are smoother), unfortunately, one of the difficulties in applying eDSC/hDSC for non-adiabatic dynamics simulation is that the sequential quadratic programming (SQP) implementation, which is a general algorithm for constrained optimization problem,  described in Ref. \citenum{Qiu:2024:dsc} can still be computationally demanding for large systems.

To address this issue, in this paper we propose a new algorithm for constructing eDSC/hDSC wavefunctions which provides a dramatic improvement in convergence speed. By separating the full constrained minimization problem into an un-constrained self-consistent-field (SCF) problem and an constrained non-self-consistent-field (nSCF) minimization problem, we have significantly reduced the computational cost of a eDSC/hDSC calculation, such that during the sampling for studying dynamics, the cost is reduced by a factor of 8-20 compared with our original (SQP) method, and now the total cost of eDSC/hDSC is approximately twice the cost of a Restricted Open-shell Hartree Fock (ROHF) calculation.

This paper is organized as follows. In Sec. \ref{sec:review}, we briefly review the eDSC/hDSC method. In Sec. \ref{sec:scf_diis_err}, we derive the SCF equation and the DIIS error vector (which is used to converge SCF within a DIIS algorithm) for the eDSC method. In Sec. \ref{sec:nscf_sqp}, we demonstrate how to incorporate the constrained minimization problem within nSCF steps and then solve the problem with a SQP algorithm. In Sec. \ref{sec:algorithm}, we summarize our full algorithm. In Sec. \ref{sec:hdsc}, we provide the analogous algorithm for hDSC. Alas, as in all CASSCF calculations, convergence to a saddle point can be  a problem, but in Sec. \ref{sec:saddle}, we provide a method to avoid a class of saddle points that is common in eDSC/hDSC.  Having described the algorithm comprehensively, in Sec. \ref{sec:result}, we provide numerical results to demonstrate that this new algorithm is significantly faster than applying the SQP algorithm directly. In Sec. \ref{sec:discussion}, we compare the computational cost of this algorithm to  a ROHF calculation and we also discuss the cost of nSCF steps. In Sec. \ref{sec:conclusion}, we conclude our findings.

\section{Method}
\subsection{Review of eDSC/hDSC}\label{sec:review}
As mentioned in Ref. \citenum{Qiu:2024:dsc}, the target function to minimize for both eDSC (where we excite an electron) and hDSC (where we excite a hole) can be written as
\begin{align}
    E_{\rm tot} &= w_1E_1 + w_2E_2\label{eq:e_tot}
\end{align}
where $E_1$ and $E_2$ are energies of two different restricted-open-shell electron configurations that both have a single unpaired particle (electron or hole). 
%with one being the one-electron (or hole) excitation of the other, i.e., 
Mathematically, for eDSC, the two configurations are
\begin{align}
    \ket{\Psi_1} &= \ket{1\bar{1},2\bar{2},...N\bar{N},N+1}\label{eq:config_e_1}\\
    \ket{\Psi_2} &= \ket{1\bar{1},2\bar{2},...N\bar{N},N+2}\label{eq:config_e_2}
\end{align}
and for hDSC, the two configurations are
\begin{align}
    \ket{\Psi_1} &= \ket{1\bar{1},2\bar{2},...,N\bar{N},N+1}\\
    \ket{\Psi_2} &= \ket{1\bar{1},2\bar{2},...,N,N+1\overline{N+1}}
\end{align}
$w_1$ and $w_2$ are dynamical weights that depend on a ``temperature" parameter $T$ and the energy difference $\Delta E = E_2 - E_1$, where we assume $E_2 > E_1$. The formula we proposed is as follows:
\begin{align}
    w_1(\Delta E) &= 1 - \frac{1-e^{-\Delta E/T}}{2\Delta E/T}\label{eq:w1}\\
    w_2(\Delta E) &= \frac{1-e^{-\Delta E/T}}{2\Delta E/T}\label{eq:w2}
\end{align}
The differential of Eq. \ref{eq:e_tot} is
\begin{align}
    dE_{\rm tot} &= w_1'dE_1 + w_2'dE_2
\end{align}
where
\begin{align}
    w_1' &= (1-\frac{1}{2}e^{-\Delta E/T})\label{eq:w1'}\\
    w_2' &= \frac{1}{2}e^{-\Delta E/T}\label{eq:w2'}
\end{align}
Note that $w_1'\neq \pp{w_1}{\Delta E}$. To avoid local excitations, a constraint is imposed such that the vector space spanned by active orbitals (i.e., orbitals $N+1$ and $N+2$ for eDSC and orbitals $N$ and $N+1$ for hDSC) projects equally onto the vector space spanned by orbitals from the donor (or left) and acceptor (or right) molecular fragments, i.e.,
\begin{align}
    {\rm Tr}(\hat{P}_L\hat{P}_{\rm active} - \hat{P}_R\hat{P}_{\rm active}) = 0\label{eq:q_formal}.
\end{align}
One way to construct $\hat{P}_L$ (and $\hat{P}_R$) is to orthonormalize all atomic orbitals (AO) from the corresponding fragment and make a projection operator from these orthonormal orbitals. As a reference, the expression for $\hat{P}_L$ and $\hat{P}_R$ in an AO basis is given in Sec. II.D.2 of Ref. \citenum{Qiu:2024:dsc}. Now,
using a Lagrangian multiplier, one can write the Lagrangian formally as
\begin{align}
    \mathcal{L} = w_1E_1 + w_2E_2 - \lambda {\rm Tr}(\hat{P}_L\hat{P}_{\rm active} - \hat{P}_R\hat{P}_{\rm active}) \label{eq:lag_formal}
\end{align}
In Ref. \citenum{Qiu:2024:dsc}, we applied  sequential quadratic programming (SQP) as a general protocol for solving this constrained minimization problem. Below we will introduce a algorithm (in the spirit of Ref. \citenum{Yanai2009})  that significantly reduces the computational cost. Without loss of generality, we will work with eDSC as an example for the next two sections (and point out the procedural differences  between eDSC and hDSC). Hereafter, we emphasize that all quantities will be written in a fixed molecular orbital (MO) basis to avoid dealing with any non-orthogonality issues that arise in an AO basis; all matrices are written in {\bf bold} font.

\subsection{SCF equations and the DIIS error vector}\label{sec:scf_diis_err}
For the case of eDSC, let $N_d$ represent the number of doubly occupied orbitals, and let $\bm{P}_0$ represent one half of the density matrix of the doubly-occupied orbitals (which is also the projection operator for the vector space spanned by the doubly-occupied orbitals). Let $\bm{P}_1$ and $\bm{P}_2$ represent the density matrix of the two singly-occupied orbitals, respectively (i.e., orbitals $N_d+1$ and $N_d+2$). Let $\bm{Q}$ be the difference between $\hat{P}_L$ and $\hat{P}_R$, i.e.,
\begin{align}
    \bm{Q} = \hat{P}_L - \hat{P}_R.\label{eq:Q}
\end{align}
By using the energy expression for an unrestricted Hartree Fock (UHF) calculation, one can evaluate the energies of the two relevant configurations $E_1, E_2$ to be 
\begin{align}
    E_1 &= \frac{1}{2}{\rm Tr}\left[(\bm{H}_0+\bm{F}_\alpha^1)\bm{P}_0\right]+\frac{1}{2}{\rm Tr}\left[(\bm{H}_0+\bm{F}_\beta^1)(\bm{P}_0+\bm{P}_1)\right]\label{eq:E1_eDSC}\\
    E_2 &= \frac{1}{2}{\rm Tr}\left[(\bm{H}_0+\bm{F}_\alpha^2)\bm{P}_0\right]+\frac{1}{2}{\rm Tr}\left[(\bm{H}_0+\bm{F}_\beta^2)(\bm{P}_0+\bm{P}_2)\right]\label{eq:E2_eDSC}
\end{align}
where $\bm{H}_0$ is the one-electron Hamiltonian; $\bm{F}_\alpha^1$, $\bm{F}_\beta^1$ are the spin up and spin down Fock matrices built from spin densities $\bm{P}_\alpha = \bm{P}_0$ and  $\bm{P}_\beta = \bm{P}_0+\bm{P}_1$, respectively; and $\bm{F}_\alpha^2$, $\bm{F}_\beta^2$ are the spin up and spin down Fock matrices built from spin densities $\bm{P}_\alpha = \bm{P}_0$, $\bm{P}_\beta = \bm{P}_0+\bm{P}_2$, respectively. Minimizing the Lagrangian in Eq. \ref{eq:lag_formal} is equivalent to minimizing the energy in Eq. \ref{eq:e_tot} w.r.t. $\bm{P}_0$, $\bm{P}_1$, and $\bm{P}_2$ given the constraints that
\begin{align}
    \bm{P}_0^2 &= \bm{P}_0\label{eq:constraint_iden1}\\
    \bm{P}_1^2 &= \bm{P}_1\\
    \bm{P}_2^2 &= \bm{P}_2\label{eq:constraint_iden2}\\
    \bm{P}_0\bm{P}_1 &= \bm{P}_0\bm{P}_2 = \bm{P}_1\bm{P}_2 = \bm{0}\label{eq:constraint_orth}\\
    {\rm Tr}\left[\bm{Q}(\bm{P}_1+\bm{P}_2)\right] &= 0\label{eq:constraint_q}
\end{align}
where $\bm{Q}$ is a given frozen matrix for each nuclear geometry. Note that if using the expression for $\bm{Q}$ (i.e., $\hat{P}_L$ and $\hat{P}_R$) in Sec. II.D.2 of Ref. \citenum{Qiu:2024:dsc} (which is expressed in an AO basis), a further basis transformation is applied to convert it into the fixed MO basis. In words, Eqs. \ref{eq:constraint_iden1}-\ref{eq:constraint_iden2} dictate that the density matrices (which are also projection matrices) are idempotent, Eq. \ref{eq:constraint_orth} requires that all density matrices be orthogonal to each other, and Eq. \ref{eq:constraint_q} is equivalent to Eq. \ref{eq:q_formal}, i.e. the constraint which forbids local excitations.  
At this point, one may rewrite the Lagrangian in Eq. \ref{eq:lag_formal}  as
\begin{align}
    \mathcal{L} = &w_1E_1+w_2E_2 - {\rm Tr}[\bm{\epsilon}_0(\bm{P}_0^2-\bm{P}_0)]-{\rm Tr}[\bm{\epsilon}_1(\bm{P}_1^2-\bm{P}_1)]-{\rm Tr}[\bm{\epsilon}_2(\bm{P}_2^2-\bm{P}_2)]\nonumber\\ 
    -&{\rm Tr}[\bm{\sigma}_1(\bm{P}_0\bm{P}_1+\bm{P}_1\bm{P}_0)]-{\rm Tr}[\bm{\sigma}_2(\bm{P}_0\bm{P}_2+\bm{P}_2\bm{P}_0)]-{\rm Tr}[\bm{\sigma}_0(\bm{P}_1\bm{P}_2+\bm{P}_2\bm{P}_1)]\nonumber\\ 
    -&\lambda{\rm Tr}[\bm{Q}(\bm{P}_1+\bm{P}_2)]\label{eq:lag_p_formal}
\end{align}
where $\bm{\epsilon}_0,\bm{\epsilon}_1,\bm{\epsilon}_2,\bm{\sigma}_1,\bm{\sigma}_2,\bm{\sigma}_0,$ and $\lambda$ are all Lagrangian multipliers. Note that we have symmetrized the constraints in Eq. \ref{eq:constraint_orth} such that not only the matrices $\bm{\epsilon}_0,\bm{\epsilon}_1,\bm{\epsilon}_2$ but also $\bm{\sigma}_1,\bm{\sigma}_2,\bm{\sigma}_0$ are symmetric. One can verify that the symmetrized constraints are equivalent to the original constraints in Eq. \ref{eq:constraint_orth} by noting that, for example,
\begin{align}
    &\bm{P}_0\bm{P}_1+\bm{P}_1\bm{P}_0 = \bm{0}\\
    \Rightarrow&\bm{P}_0\bm{P}_1+\bm{P}_1\bm{P}_0\bm{P}_1 = \bm{0}\\
    \Rightarrow&\bm{P}_1\bm{P}_0+\bm{P}_1\bm{P}_0\bm{P}_1 = \bm{0}\\
    \Rightarrow&\bm{P}_0\bm{P}_1=\bm{P}_1\bm{P}_0 = \bm{0}.
\end{align}

Now, to find a eDSC solution and solve the constrained minimization problem in Eq. \ref{eq:lag_p_formal}, the gradient of $\mathcal{L}$ w.r.t. the density matrices must be zero:
\begin{align}
    \grad{P_0}\mathcal{L}&=w_1'(\bm{F}_\alpha^1+\bm{F}_\beta^1)+w_2'(\bm{F}_\alpha^2+\bm{F}_\beta^2)\nonumber\\
    &-\bm{P}_0\bm{\epsilon}_0-\bm{\epsilon}_0\bm{P}_0+\bm{\epsilon}_0-\bm{\sigma}_1\bm{P}_1-\bm{P}_1\bm{\sigma}_1-\bm{\sigma}_2\bm{P}_2-\bm{P}_2\bm{\sigma}_2=\bm{0}\label{eq:L_grad_p1}\\
    \grad{P_1}\mathcal{L}&=w_1'\bm{F}_\beta^1-\bm{P}_1\bm{\epsilon}_1-\bm{\epsilon}_1\bm{P}_1+\bm{\epsilon}_1-\bm{P}_0\bm{\sigma}_1-\bm{\sigma}_1\bm{P}_0-\bm{\sigma}_0\bm{P}_2-\bm{P}_2\bm{\sigma}_0-\lambda \bm{Q}=\bm{0}\\
    \grad{P_2}\mathcal{L}&=w_2'\bm{F}_\beta^2-\bm{P}_2\bm{\epsilon}_2-\bm{\epsilon}_2\bm{P}_2+\bm{\epsilon}_2-\bm{P}_0\bm{\sigma}_2-\bm{\sigma}_2\bm{P}_0-\bm{\sigma}_0\bm{P}_1-\bm{P}_1\bm{\sigma}_0-\lambda \bm{Q}=\bm{0}\label{eq:L_grad_p2}
\end{align}
Multiplying Eqs. \ref{eq:L_grad_p1}-\ref{eq:L_grad_p2}  on the right by $\bm{P}_0,\bm{P}_1,\bm{P}_2$, respectively, yields
\begin{align}
    (w_1'\bm{F}_\alpha^1+w_2'\bm{F}_\alpha^2+w_1'\bm{F}_\beta^1+w_2'\bm{F}_\beta^2)\bm{P}_0 &=\bm{P}_0\bm{\epsilon}_0\bm{P}_0 +\bm{P}_1\bm{\sigma}_1\bm{P}_0+\bm{P}_2\bm{\sigma}_2\bm{P}_0\label{eq:dP1}\\
    (w_1'\bm{F}_\beta^1-\lambda \bm{Q})\bm{P}_1&=\bm{P}_0\bm{\sigma}_1\bm{P}_1 +\bm{P}_1\bm{\epsilon}_1\bm{P}_1+\bm{P}_2\bm{\sigma}_0\bm{P}_1\label{eq:dP2}\\
    (w_2'\bm{F}_\beta^2-\lambda \bm{Q})\bm{P}_2&=\bm{P}_0\bm{\sigma}_2\bm{P}_2 +\bm{P}_1\bm{\sigma}_0\bm{P}_2+\bm{P}_2\bm{\epsilon}_2\bm{P}_2\label{eq:dP3}
\end{align}
One may use a shorthand for the weighted Fock matrices:
\begin{align}
    \bm{M}_0&=w_1'\bm{F}_\alpha^1+w_2'\bm{F}_\alpha^2+w_1'\bm{F}_\beta^1+w_2'\bm{F}_\beta^2\label{eq:M0}\\
    \bm{M}_1&=w_1'\bm{F}_\beta^1\\
    \bm{M}_2&=w_2'\bm{F}_\beta^2\label{eq:M2}
\end{align}
and Eqs. \ref{eq:dP1}-\ref{eq:dP3} become
\begin{align}
    \bm{M}_0\bm{P}_0 &=\bm{P}_0\bm{\epsilon}_0\bm{P}_0 +\bm{P}_1\bm{\sigma}_1\bm{P}_0+\bm{P}_2\bm{\sigma}_2\bm{P}_0\label{eq:dP1_M}\\
    (\bm{M}_1-\lambda \bm{Q})\bm{P}_1&=\bm{P}_0\bm{\sigma}_1\bm{P}_1 +\bm{P}_1\bm{\epsilon}_1\bm{P}_1+\bm{P}_2\bm{\sigma}_0\bm{P}_1\label{eq:dP2_M}\\
    (\bm{M}_2-\lambda \bm{Q})\bm{P}_2&=\bm{P}_0\bm{\sigma}_2\bm{P}_2 +\bm{P}_1\bm{\sigma}_0\bm{P}_2+\bm{P}_2\bm{\epsilon}_2\bm{P}_2.\label{eq:dP3_M}
\end{align}
Since $w_1',w_2',\bm{F}_\alpha^1,\bm{F}_\alpha^2,\bm{F}_\beta^1,\bm{F}_\beta^2$ all depend on $\bm{P}_0,\bm{P}_1,\bm{P}_2$ (and so do $\bm{M}_0$, $\bm{M}_1$, and $\bm{M}_2$ in Eqs. \ref{eq:dP1_M}-\ref{eq:dP3_M}), Eqs. \ref{eq:dP1}-\ref{eq:dP3} are effectively  ``SCF'' equations. Although solving these equations directly is difficult, an easier approach is to separate the procedure into two parts. 
To better appreciate our two-pronged approach, note that 
%all of the matrices in Eqs. \ref{eq:dP1}-\ref{eq:dP3} are symmetric and, 
if one sums up the R.H.S. of Eq. \ref{eq:dP1_M} + Eq. \ref{eq:dP2_M} + Eq. \ref{eq:dP3_M}, the result is symmetric, so that
\begin{align}
    [\bm{M}_0,\bm{P}_0]+[\bm{M}_1-\lambda \bm{Q},\bm{P}_1]+[\bm{M}_2-\lambda \bm{Q},\bm{P}_2] = \bm{0}\label{eq:diis_err_0}
\end{align}
where $[\bm{A},\bm{B}]=\bm{AB}-\bm{BA}$ is the commutator. Thus, Eq. \ref{eq:diis_err_0} suggests that one can use
\begin{align}
  \bm{V}_{\rm DIIS} =   [\bm{M}_0,\bm{P}_0]+[\bm{M}_1-\lambda \bm{Q},\bm{P}_1]+[\bm{M}_2-\lambda \bm{Q},\bm{P}_2]\label{eq:diis_err}
\end{align}
as an error vector indicating how far $\bm{P}_0,\bm{P}_1,\bm{P}_2$ are away from the desired solution.
Thus, our approach is as follows:
\begin{enumerate}
\item First, we will solve Eqs. \ref{eq:dP1}-\ref{eq:dP3} for $\bm{P}_0,\bm{P}_1,\bm{P}_2$ given fixed quantities of $w_1',w_2',\bm{F}_\alpha^1,\bm{F}_\alpha^2,\bm{F}_\beta^1,\bm{F}_\beta^2$ (or equivalently, $\bm{M}_0,\bm{M}_1,\bm{M}_2$).  Henceforward, we will refer to solving Eqs. \ref{eq:dP1}-\ref{eq:dP3} with fixed $\bm{M}_0,\bm{M}_1,\bm{M}_2$ as the non-SCF (or nSCF) part of the problem.
\item Second, using the density matrices (i.e., $\bm{P}_0,\bm{P}_1,\bm{P}_2$) just computed and the error vector in Eq. \ref{eq:diis_err}, we
will establish a DIIS procedure\cite{pulay:1982} for solving Eqs. \ref{eq:dP1}-\ref{eq:dP3} self-consistently.  
\end{enumerate}

\subsection{nSCF steps and SQP}\label{sec:nscf_sqp}
For fixed $\bm{M}_0,\bm{M}_1,\bm{M}_2$, satisfying Eqs. \ref{eq:dP1}-\ref{eq:dP3} is equivalent to minimizing an auxiliary function of the form:
\begin{align}
    E_{\rm aux} = {\rm Tr}\left[\bm{M}_0\bm{P}_0\right]+{\rm Tr}\left[\bm{M}_1\bm{P}_1\right]+{\rm Tr}\left[\bm{M}_2\bm{P}_2\right]\label{eq:sqp_E}
\end{align}
given the constraints in Eqs. \ref{eq:constraint_iden1}-\ref{eq:constraint_q}. Now, one can avoid the orthonormal constraints in Eqs. \ref{eq:constraint_iden1}-\ref{eq:constraint_orth} by defining a unitary matrix $\bm{C}$ and occupation matrices $\bm{K}_0,\bm{K}_1,\bm{K}_2$ such that
\begin{align}
    \bm{P}_0 &= \bm{C}\bm{K}_0\bm{C}^\top\label{eq:P0}\\
    \bm{P}_1 &= \bm{C}\bm{K}_1\bm{C}^\top\\
    \bm{P}_2 &= \bm{C}\bm{K}_2\bm{C}^\top\label{eq:P2}
\end{align}
where $\bm{K}_0,\bm{K}_1,\bm{K}_2$ are diagonal matrices with diagonal elements equal to 0 or 1 (depending on whether the orbitals are occupied in the relevant density matrix). Moreover, as far as minimizing $E_{\rm aux}$, the unitary $\bm{C}$ matrix can be parameterized by an anti-symmetric matrix $\bm{A}$, i.e.,
\begin{align}
    \bm{C} = \bm{C}_0e^{\bm{A}}
\end{align}
such that the constraints in Eqs. \ref{eq:constraint_iden1}-\ref{eq:constraint_orth} are satisfied automatically.

Thus, in the end, the only unavoidable constraint is Eq. \ref{eq:constraint_q}, which we may name as
\begin{align}
    G = {\rm Tr}\left[\bm{Q}(\bm{P}_1+\bm{P}_2)\right] = 0.\label{eq:sqp_G}
\end{align}
Minimizing $E_{\rm aux}$ with this constraint is equivalent to solving a Lagrangian of the form:
\begin{align}
    \mathcal{L}_{\rm aux}(\bm{A}) = &{\rm Tr}\left[e^{-\bm{A}}\bm{C}_0^\top\bm{M}_0\bm{C}_0e^{\bm{A}}\bm{K}_0\right]+{\rm Tr}\left[e^{-\bm{A}}\bm{C}_0^\top\bm{M}_1\bm{C}_0e^{\bm{A}}\bm{K}_1\right]+{\rm Tr}\left[e^{-\bm{A}}\bm{C}_0^\top\bm{M}_2\bm{C}_0e^{\bm{A}}\bm{K}_2\right]\nonumber\\
    -&\lambda{\rm Tr}\left[e^{-\bm{A}}\bm{C}_0^\top\bm{Q}\bm{C}_0e^{\bm{A}}(\bm{K}_1+\bm{K}_2)\right]
\end{align}
One can solve this constrained minimization problem using the SQP\cite{Nocedal2006} algorithm, which requires only the gradients of the target function (Eq. \ref{eq:sqp_E}) and the constraint (Eq. \ref{eq:sqp_G}) w.r.t. $\bm{A}$. The expressions for these gradients at $\bm{A}=\bm{0}$ can be found easily by using Baker–Campbell–Hausdorff formula, and the results are (noting that $\bm{A}$ is anti-symmetric)
\begin{align}
    \left.\pp{E_{\rm aux}}{A_{pq}}\right|_{\bm{A}=\bm{0}} = 2\left([\tilde{\bm{M}}_0,\bm{K}_0]+[\tilde{\bm{M}}_1,\bm{K}_1]+[\tilde{\bm{M}}_2,\bm{K}_2]\right)_{pq}\label{eq:dEdA}
\end{align}
for $q>p$, where
\begin{align}
    \tilde{\bm{M}}_0 &= \bm{C}_0^\top \bm{M}_0 \bm{C}_0\\
    \tilde{\bm{M}}_1 &= \bm{C}_0^\top \bm{M}_1 \bm{C}_0\\
    \tilde{\bm{M}}_2 &= \bm{C}_0^\top \bm{M}_2 \bm{C}_0
\end{align}
and
\begin{align}
    \left.\pp{G}{A_{pq}}\right|_{\bm{A}=\bm{0}} = 2\left([\tilde{\bm{Q}},\bm{K}_1+\bm{K}_2]\right)_{pq}\label{eq:dQdA}
\end{align}
for $q>p$, where
\begin{align}
    \tilde{\bm{Q}} = \bm{C}_0^\top \bm{Q} \bm{C}_0
\end{align}
Because Eqs \ref{eq:dEdA} and \ref{eq:dQdA} are valid only at $\bm{A}=\bm{0}$, during an SQP minimization, $\bm{C}_0$ must be updated in each iteration. 

Finally, for practical optimization problems, it is often desirable to use the diagonal of the Hessian matrix as a preconditioner. These elements can be derived  using a Baker–Campbell–Hausdorff formula, and one finds
\begin{align}
    W_{pq}\equiv\left.\pp{^2E_{\rm aux}}{A_{pq}^2}\right|_{\bm{A}=\bm{0}} &= 2\left((K_{0,pp}-K_{0,qq})(\tilde{M}_{0,qq}-\tilde{M}_{0,pp})+ (K_{1,pp}-K_{1,qq})(\tilde{M}_{1,qq}-\tilde{M}_{1,pp})\right.\nonumber\\
    &+\left.(K_{2,pp}-K_{2,qq})(\tilde{M}_{2,qq}-\tilde{M}_{2,pp})\right)\label{eq:w_pq}
\end{align}
Mathematically, this preconditioing process works as follows.  Instead of using Eq. \ref{eq:dEdA} and \ref{eq:dQdA} as the gradients of the target function and the constraint used in SQP algorithm,
we define these gradients  ($df_{pq}$ and $dg_{pq}$, respectively)   as
\begin{align}
    df_{pq} &= 2\left.\left([\tilde{\bm{M}}_0,\bm{K}_0]+[\tilde{\bm{M}}_1,\bm{K}_1]+[\tilde{\bm{M}}_2,\bm{K}_2]\right)_{pq} \right/ |W_{pq}|\label{eq:sqp_df}\\
    dg_{pq} &= 2\left.\left([\tilde{\bm{Q}},\bm{K}_1+\bm{K}_2]\right)_{pq} \right/ |W_{pq}|.\label{eq:sqp_dg}
\end{align}
where $W_{pq}$ is defined in Eq. \ref{eq:w_pq}. 

Finally, let us briefly review the SQP procedure for minimization given gradients $df_{pq}$ and $dg_{pq}$; the readers may refer to Ref. \citenum{Nocedal2006} for more details. The elements $df_{pq}$ and $dg_{pq}$ ($q>p$) may be regarded as the elements of two vectors, $\bm{df}$ and $\bm{dg}$, respectively. At each iteration, our goal is to find an updated $\lambda$ value and a new $\bm{A}$ matrix so as to rotate the orbitals. Here we use $\bm{dA}$ to represent the vector constituted by the matrix element $A_{pq}$, $q>p$. The SQP procedure breaks $\bm{dA}$ into two components, $\bm{dA}_\parallel$ and $\bm{dA}_\bot$, such that $\bm{dA}_\parallel$ is parallel to $\bm{dg}$ and $\bm{dA}_\bot$ is perpendicular to $\bm{dg}$. At each iteration, the algorithm first calculates an updated $\lambda$
\begin{align}
    \lambda = \left(\bm{df}^\top\bm{dg}\right) / \left(\bm{dg}^\top\bm{dg}\right)\label{eq:new_lambda}
\end{align}
as well as $\bm{dA}_\parallel$:
\begin{align}
    \bm{dA}_\parallel = -\left({\rm Tr}\left[\tilde{\bm{Q}}(\bm{K}_1+\bm{K}_2)\right] / \left(\bm{dg}^\top\bm{dg}\right)\right) \bm{dg}.
\end{align}
$\bm{dA}_\bot$ is the solution  to the equation
\begin{align}
    \bm{H}_\bot\bm{dA}_\bot = \bm{y} \implies \bm{dA}_\bot = \bm{B}\bm{y}\label{eq:sqp_perpendicular}
\end{align}
Here $\bm{H}_\bot$ is the Hessian in the space perpendicular to the gradient of the constraint ($\bm{dg}$), $\bm{B}$ is the inverse of $\bm{H}_\bot$, and $\bm{y}$ is the difference between Lagrangian calculated at current iteration and previous iteration in the space perpendicular to $\bm{dg}$, i.e.,
\begin{align}
    \bm{y} = \left(1-\frac{\bm{dg}\bm{dg}^\top}{\bm{dg}^\top\bm{dg}}\right)\left(\Delta \bm{df} - \lambda \Delta \bm{dg}\right)
\end{align}
given that $\Delta \bm{df}$ (and $\Delta \bm{dg}$) are differences between $\bm{df}$ (and $\bm{dg}$) calculated at current iteration and previous iteration. Note that $\lambda$ is taken from Eq. \ref{eq:new_lambda} and does not include quantities calculated at a previous iteration. 

For large systems, it becomes very time consuming to solve Eq. \ref{eq:sqp_perpendicular} directly or even with the BFGS algorithm. To that end,  in practice,  a limited-memory BFGS (l-BFGS) algorithm is applied that solves Eq. \ref{eq:sqp_perpendicular} iteratively given the vectors $\bm{dA}_\bot$ and $\bm{y}$ from previous iterations.  Once the new $\bm{dA}_\bot$ is achieved, $\bm{dA} = \bm{dA}_\parallel+\bm{dA}_\bot$ provides the direction to rotate the orbitals. In principle one needs a line search step to determine the step size.  However, we find that, as long as the preconditioner from Eqs. \ref{eq:sqp_df} and \ref{eq:sqp_dg} is applied, and the calculation starts from an initial guess that is close to the solution, one can safely choose the step size to be 1. Note that, in practice, we can find an appropriate initial guess by either  reading a solution from a similar nuclear geometry or reading a solution from previous SCF step or relaxing the system for a few steps by setting the memory length of l-BFGS to one or two with small step size (say, a 0.2 step size). 

\subsection{Full procedure}\label{sec:algorithm}
Let us now summarize the full procedure of our algorithm.
\begin{enumerate}[{\bf Step 1:}]
    \item Choose a set of MO as the basis, calculate $\bm{Q}$ in this basis (Eq. \ref{eq:Q}, also see Sec. II.D.2 of Ref. \citenum{Qiu:2024:dsc}), choose an initial guess $\bm{C}_{\rm ini}$  for $\bm{C}$ and an initial guess for $\lambda$. Choose a DIIS threshold $T_{D}$. One may use the fixed MO basis as the initial guess, in which case $\bm{C}_{\rm ini}$ is an identity matrix.
    \item Build density matrices $\bm{P}_0,\bm{P}_1,\bm{P}_2$ from $\bm{C}$ using Eqs. \ref{eq:P0}-\ref{eq:P2}.
    \item Build $w_1',w_2',\bm{F}_\alpha^1,\bm{F}_\alpha^2,\bm{F}_\beta^1,\bm{F}_\beta^2$ from $\bm{P}_0,\bm{P}_1,\bm{P}_2$ (Eqs. \ref{eq:w1'}-\ref{eq:w2'}, \ref{eq:E1_eDSC}-\ref{eq:E2_eDSC}), then build $\bm{M}_0,\bm{M}_1,\bm{M}_2$ from Eqs. \ref{eq:M0}-\ref{eq:M2}.
    \item Compute the initial DIIS error vector $\bm{V}^0$ using Eq. \ref{eq:diis_err}, i.e.,
    \begin{align}
        \bm{V}^0 = [\bm{M}_0,\bm{P}_0]+[\bm{M}_1-\lambda \bm{Q},\bm{P}_1]+[\bm{M}_2-\lambda \bm{Q},\bm{P}_2]
    \end{align}
    \item If norm $(\bm{V}^n) < T_D$, jump to {\bf Step} \ref{step:diis_stop}. Otherwise, set the initial DIIS variable $\bm{A}^0$ to $\bm{0}$.
    \item Solve the nSCF problem for the new $\bm{C}$ and $\lambda$ using SQP with Eqs. \ref{eq:sqp_df}-\ref{eq:sqp_dg} given $\bm{M}_0,\bm{M}_1,\bm{M}_2$, and $\bm{Q}$. The nSCF threshold for the norm of the gradient and the constraint is set to norm $(\bm{V}^0)/100$.
    \item (DIIS iteration starts. Let the iteration counter $n$ starts from 1.) Build density matrices $\bm{P}_0,\bm{P}_1,\bm{P}_2$ from new $\bm{C}$ using Eqs. \ref{eq:P0}-\ref{eq:P2}.\label{step:diis_start}
    \item Build $w_1',w_2',\bm{F}_\alpha^1,\bm{F}_\alpha^2,\bm{F}_\beta^1,\bm{F}_\beta^2$ from $\bm{P}_0,\bm{P}_1,\bm{P}_2$ (Eqs. \ref{eq:w1'}-\ref{eq:w2'}, \ref{eq:E1_eDSC}-\ref{eq:E2_eDSC}), then build $\bm{M}_0,\bm{M}_1,\bm{M}_2$ from Eqs. \ref{eq:M0}-\ref{eq:M2}.\label{step:build_fock}
    \item Calculate the DIIS error vector at iteration $n$:
    \begin{align}
        \bm{V}^n = [\bm{M}_0,\bm{P}_0]+[\bm{M}_1-\lambda \bm{Q},\bm{P}_1]+[\bm{M}_2-\lambda \bm{Q},\bm{P}_2]
    \end{align}
    \item Solve the nSCF problem for new $\bm{C}$ and $\lambda$ using SQP with Eqs. \ref{eq:sqp_df}-\ref{eq:sqp_dg} given $\bm{M}_0,\bm{M}_1,\bm{M}_2$, and $\bm{Q}$. The nSCF threshold for the norm of the gradient and the constraint is set to norm $(\bm{V}^n)/100$.\label{step:nscf}
    \item Calculate the DIIS variable at iteration $n$:
    \begin{align}
        \bm{A}^n = \log\left(\bm{C}_{\rm ini}^\top\bm{C}\right)
    \end{align}
    \item If norm $(\bm{V}^n) < T_D$, exit the iteration and jump to {\bf Step} \ref{step:diis_stop}. Otherwise, use DIIS algorithm to calculate new $\bm{C}$. Specifically, solve the matrix equation
    \begin{align}
        \begin{bmatrix}
            B_{00} &B_{01} &B_{02} &\cdots &B_{0n} &1\\
            B_{10} &B_{11} &B_{12} &\cdots &B_{1n} &1\\
            B_{20} &B_{21} &B_{22} &\cdots &B_{2n} &1\\
            \vdots &\vdots &\vdots &\ddots &\vdots &\vdots\\
            B_{n0} &B_{n1} &B_{n2} &\cdots &B_{nn} &1\\
            1      &1      &1      &\cdots &1      &0
        \end{bmatrix}
        \begin{bmatrix}
            c_0\\ c_1\\ c_2\\ \vdots\\ c_n\\ \zeta
        \end{bmatrix}
        =
        \begin{bmatrix}
            0\\ 0\\ 0\\ \vdots\\ 0\\ 1
        \end{bmatrix}
    \end{align}
    for $c_0,c_1,c_2,\cdots,c_n$, where $B_{ij} = {\rm Tr}\left[\bm{V}^{i\top}\bm{V}^j\right]$, calculate $\Delta \bm{A}$ as 
    \begin{align}
        \Delta \bm{A} = \sum_{i=0}^n c_i\bm{A}^i
    \end{align}
    and set $\bm{C}$ to be $\bm{C}_{\rm ini}e^{\Delta \bm{A}}$.
    \item Return to {\bf Step} \ref{step:diis_start}.
    \item DIIS convergence achieved.\label{step:diis_stop}
\end{enumerate}
\subsection{Case of hDSC}\label{sec:hdsc}
In Secs. \ref{sec:scf_diis_err}-\ref{sec:algorithm} above, we have discussed in detail our new approach for constructing an eDSC solution. It is important to note that a very anologous procedure also applies to hDSC wavefunction, except that some quantities need to be redefined. In this section, we will now show how to modify such  definitions correctly.  

To begin  with, in a manner analogous to Eqs. \ref{eq:E1_eDSC}-\ref{eq:E2_eDSC}, the energy expressions for hDSC can be written as
\begin{align}
    E_1 &= \frac{1}{2}{\rm Tr}\left[(\bm{H}_0+\bm{F}_\alpha^1)(\bm{P}_0+\bm{P}_1+\bm{P}_2)\right]+\frac{1}{2}{\rm Tr}\left[(\bm{H}_0+\bm{F}_\beta^1)(\bm{P}_0+\bm{P}_1)\right]\label{eq:E1_hDSC}\\
    E_2 &= \frac{1}{2}{\rm Tr}\left[(\bm{H}_0+\bm{F}_\alpha^2)(\bm{P}_0+\bm{P}_1+\bm{P}_2)\right]+\frac{1}{2}{\rm Tr}\left[(\bm{H}_0+\bm{F}_\beta^2)(\bm{P}_0+\bm{P}_2)\right]\label{eq:E2_hDSC}
\end{align}
where $\bm{P}_0$ sums up the contributions of all orbitals through $N_d-1$, and $\bm{P}_1$ and $\bm{P}_2$ are projections operator for orbital $N_d$ and $N_d+1$, respectively.  

Thereafter, 
\begin{itemize}
    \item $\bm{K}_0,\bm{K}_1,\bm{K}_2$ in Eqs. \ref{eq:P0}-\ref{eq:P2} must be adjusted accordingly;
    \item The matrices $\bm{F}_\alpha^1$ and $\bm{F}_\beta^1$ must be built from spin densities $\bm{P}_\alpha = \bm{P}_0+\bm{P}_1+\bm{P}_2$ and  $\bm{P}_\beta = \bm{P}_0+\bm{P}_1$;
    \item The matrices $\bm{F}_\alpha^2$ and $\bm{F}_\beta^2$ must built from spin densities  $\bm{P}_\alpha = \bm{P}_0+\bm{P}_1+\bm{P}_2$, $\bm{P}_\beta = \bm{P}_0+\bm{P}_2$. 
    \end{itemize}

    Lastly,  Eqs. \ref{eq:dP1}-\ref{eq:dP3} is replaced by
\begin{align}
    (w_1'\bm{F}_\alpha^1+w_2'\bm{F}_\alpha^2+w_1'\bm{F}_\beta^1+w_2'\bm{F}_\beta^2)\bm{P}_0 &=\bm{P}_0\bm{\epsilon}_0\bm{P}_0 +\bm{P}_1\bm{\sigma}_1\bm{P}_0+\bm{P}_2\bm{\sigma}_2\bm{P}_0\\
    (w_1'\bm{F}_\alpha^1+w_2'\bm{F}_\alpha^2+w_1'\bm{F}_\beta^1-\lambda \bm{Q})\bm{P}_1&=\bm{P}_0\bm{\sigma}_1\bm{P}_1 +\bm{P}_1\bm{\epsilon}_1\bm{P}_1+\bm{P}_2\bm{\sigma}_0\bm{P}_1\\
    (w_1'\bm{F}_\alpha^1+w_2'\bm{F}_\alpha^2+w_2'\bm{F}_\beta^2-\lambda \bm{Q})\bm{P}_2&=\bm{P}_0\bm{\sigma}_2\bm{P}_2 +\bm{P}_1\bm{\sigma}_0\bm{P}_2+\bm{P}_2\bm{\epsilon}_2\bm{P}_2
\end{align}
which implies that one must define
\begin{align}
    \bm{M}_0&=w_1'\bm{F}_\alpha^1+w_2'\bm{F}_\alpha^2+w_1'\bm{F}_\beta^1+w_2'\bm{F}_\beta^2\\
    \bm{M}_1&=w_1'\bm{F}_\alpha^1+w_2'\bm{F}_\alpha^2+w_1'\bm{F}_\beta^1\\
    \bm{M}_2&=w_1'\bm{F}_\alpha^1+w_2'\bm{F}_\alpha^2+w_2'\bm{F}_\beta^2
\end{align}
instead of Eqs. \ref{eq:M0}-\ref{eq:M2}. Given the changes above, the algorithm in Sec. \ref{sec:algorithm} can be applied directly to hDSC.

\subsection{Avoiding the saddle point}\label{sec:saddle}
For systems with very similar $D_0$ and $D_1$ energies, we have found that the algorithm described above can sometimes converge to a saddle point.  To understand why we find convergence to a saddle point,  note that, in an MO basis, the coupling between the two active orbitals is identical for $\bm{F}_\beta^1$ and $\bm{F}_\beta^2$, i.e. for eDSC
\begin{align}
    \left(\tilde{\bm{F}}_\beta^1\right)_{N_d+1,N_d+2} = \left(\tilde{\bm{F}}_\beta^2\right)_{N_d+1,N_d+2}
\end{align}
where 
\begin{align}
    \tilde{\bm{F}}_\beta^1 &= \bm{C}^\top \bm{F}_\beta^1\bm{C}\\
    \tilde{\bm{F}}_\beta^2 &= \bm{C}^\top \bm{F}_\beta^2\bm{C}\label{eq:F2_MO_basis}
\end{align}
and for hDSC
\begin{align}
    \left(\tilde{\bm{F}}_\beta^1\right)_{N_d,N_d+1} = \left(\tilde{\bm{F}}_\beta^2\right)_{N_d,N_d+1}
\end{align}
Now, if $w_1 = w_2$, one can verify that these two terms (i.e., $\left(\tilde{\bm{F}}_\beta^1\right)_{N_d+1,N_d+2}$ and $\left(\tilde{\bm{F}}_\beta^2\right)_{N_d+1,N_d+2}$ for eDSC; $\left(\tilde{\bm{F}}_\beta^1\right)_{N_d,N_d+1}$ and $\left(\tilde{\bm{F}}_\beta^2\right)_{N_d,N_d+1}$ for hDSC) in Eq. \ref{eq:diis_err} cancel with each other so that the error vector contains no information about these quantities. That  being said, for the proper minimum we are looking for, we seek a solution for which both terms
%($\left(\tilde{\bm{F}}_\beta^1\right)_{N_d+1,N_d+2}$ and $\left(\tilde{\bm{F}}_\beta^2\right)_{N_d+1,N_d+2}$ for eDSC; $\left(\tilde{\bm{F}}_\beta^1\right)_{N_d,N_d+1}$ and $\left(\tilde{\bm{F}}_\beta^2\right)_{N_d,N_d+1}$ for hDSC)
vanish independently. This dichotomy (i.e. the need for a solution with two vanishing matrix elements but an algorithm that has information only about the sum of the two matrix elements) leads to a situation whereby the system can easily fall into a saddle point, which significantly limits the convergence of the algorithm as well as the quality of the solution in an avoided crossing region. 

To overcome this issue, in principle one could perform a hessian computation and find the mode corresponding to the negative eigenvalue (and then walk in that direction before a second optimization is launched). As a much cheaper alternative, however, we have found that one can avoid such ``saddle point'' entirely simply by scaling one of the two relevant terms. Taking eDSC as an example, we have found much improved convergence by scaling
one of these two terms with a factor of minus one, e.g.,
\begin{align}
    \left(\tilde{\bm{F}}_\beta^2\right)_{N_d+1,N_d+2} \longrightarrow -\left(\tilde{\bm{F}}_\beta^2\right)_{N_d+1,N_d+2}\label{eq:scale_F}
\end{align}
Note that this rescaling needs to be performed before subsequently building $\bm{M}_0,\bm{M}_1,\bm{M}_2$,. In other words,  after computing $\bm{F}_\beta^2$ from the UHF theory, one should first perform the matrix multiplication in Eq. \ref{eq:F2_MO_basis} to generate $\tilde{\bm{F}}_\beta^2$, then apply the scaling factor of -1 to the corresponding matrix element, and finally invert Eq. \ref{eq:F2_MO_basis} to compute the updated $\bm{F}_\beta^2$. All of these steps should be done before building $\bm{M}_0,\bm{M}_1,\bm{M}_2$. This approach guarantees that the contribution from $\left(\tilde{\bm{F}}_\beta^1\right)_{N_d+1,N_d+2}$ in the DIIS error vector is $w_1+w_2 = 1$, so that this quantity must vanish at convergence. Similarly, for hDSC, we also scale the corresponding term with a factor of minus one:
\begin{align}
    \left(\tilde{\bm{F}}_\beta^2\right)_{N_d,N_d+1} \longrightarrow -\left(\tilde{\bm{F}}_\beta^2\right)_{N_d,N_d+1}\label{eq:scale_F_h}
\end{align}

As a practical matter, we have also tried other scaling factors; see Sec. \ref{sec:php} and \ref{sec:ac} for details.

\section{Numerical results}\label{sec:result}

\begin{figure*}[ht]
    \centering\includegraphics[width=0.8\textwidth]{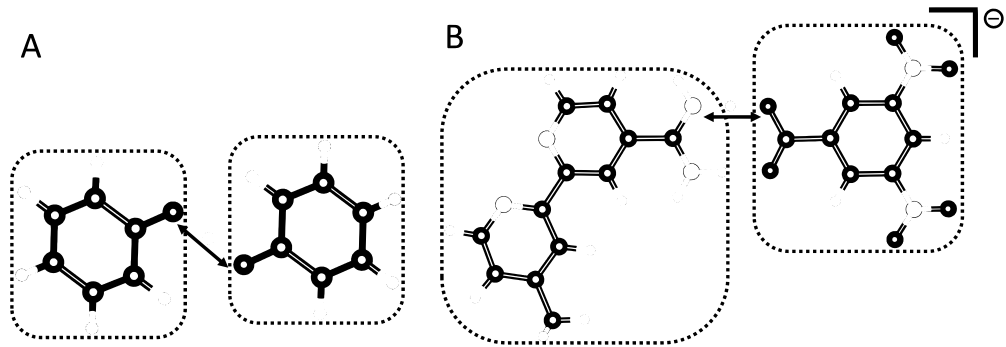}
    \caption{Geometries, partition of fragments, and reaction coordinates of studied systems. The red and blue dashed boxes illustrate the partition of left and right fragments, respectively. The black arrows indicate the reaction coordinates. (A) Phenoxyl-phenol (ph-ph), the reaction coordinate is defined by moving the hydrogen between two oxygen atoms while fixing all other atoms. (B) The amidinium-carboxylate system (am-ca). The reaction coordinate is defined by moving the hydrogen between the nitrogen and oxygen atoms while fixing all other atoms.}
    \label{fig:geo}
\end{figure*}

We have implemented the above algorithm in QChem and applied it to two systems that were studied in our previous work, namely the phenoxyl-phenol ({\bf ph-ph}) system and the amidinium-carboxylate  ({\bf am-ca}) system, where the geometries are shown in Fig. \ref{fig:geo} and the atomic coordinates are provided in the Supporting Information. As pointed out by Ref. \citenum{Qiu:2024:dsc}, hDSC is applied to {\bf ph-ph} and eDSC is applied to {\bf am-ca}. We use the same reaction coordinates as used in Ref. \citenum{Qiu:2024:dsc} and the initial guess is read from solutions at previous geometry along the reaction path. For all calculations, the gradient is converged to less than $10^{-5}$ a.u. Note that Ref. \citenum{Qiu:2024:dsc} has already demonstrated that, for these problems, eDSC/hDSC can generate smooth potential energy surfaces that express $D_0/D_1$ crossing behavior. With this fact in mind, our purpose in this section is to compare the computational cost between the method used in the previous work (Ref. \citenum{Qiu:2024:dsc}, henceforward referred to as SQP) versus the algorithm presented above in this work  (henceforward referred to as DIIS-SQP). For readers that are interested in how the temperature parameter may influence the performance of our method, please refer to the Supporting Information. We find that the performance of the algorithm is insensitive to the temperature parameter used in the dynamical weighting scheme.

\subsection{ph-ph (hDSC)}\label{sec:php}
\begin{figure*}[ht!]
    \centering
    \includegraphics[width=0.95\textwidth]{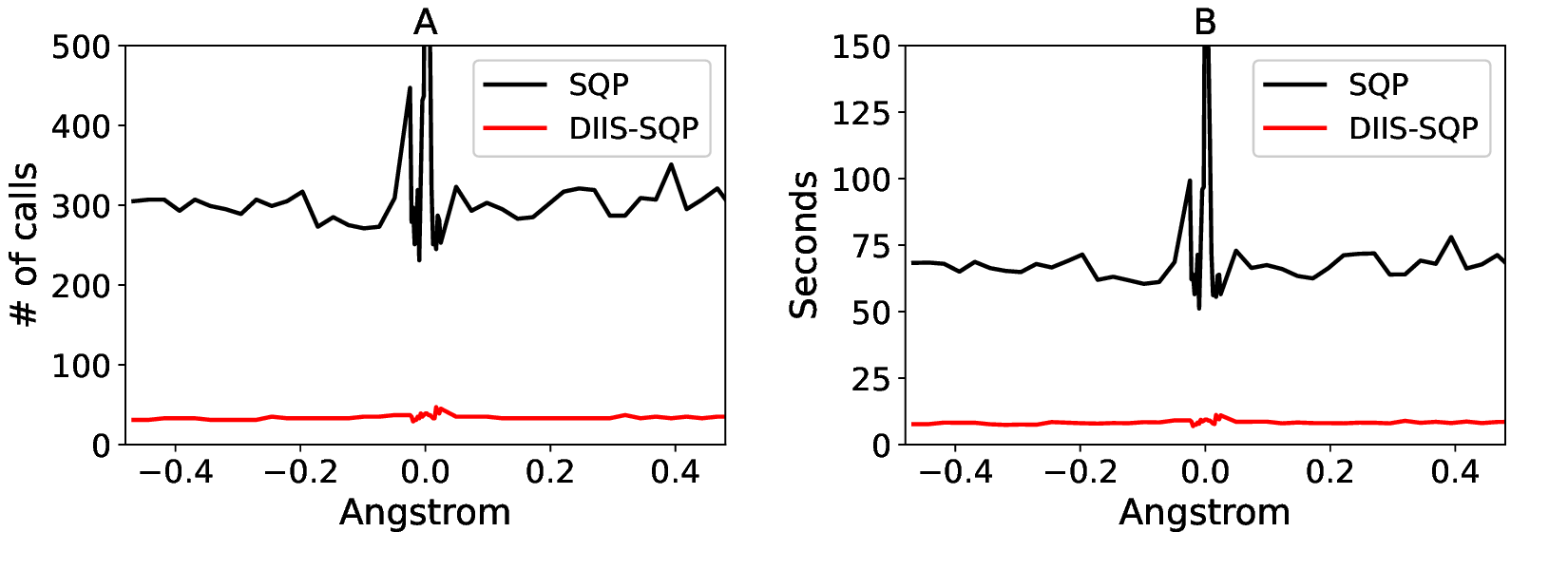}
    \caption{Computational cost of SQP and DIIS-SQP method for ph-ph system. A scaling of -1 for $\left(\tilde{\bm{F}}_\beta^2\right)_{N_d,N_d+1}$ is applied for both methods. (A) Total number of calls to build the Fock matrices, i.e., $\bm{F}_\alpha^1,\bm{F}_\alpha^2,\bm{F}_\beta^1,\bm{F}_\beta^2$. These are the most time-consuming calls. (B) Total wall time for converging the gradient to less than $10^{-5}$ a.u. when running on 48 cores. DIIS-SQP reduces the computational cost by a factor of 10 to 20 compared with only using SQP.}
    \label{fig:php_time}
\end{figure*}

In Fig. \ref{fig:php_time}, we plot the relevant computational costs for a ph-ph simulation . The horizontal axis is the displacement of the bridging hydrogen atom from the center of the adjacent two oxygen atoms. Since  building the Fock matrices (i.e., $\left(\tilde{\bm{F}}_\beta^2\right)_{N_d,N_d+1}$) dominates the total computational cost, on the vertical axis we plot the total number of Fock-build calls  both for SQP (our previous algorithm from Ref. \citenum{Qiu:2024:dsc}) and DIIS-SQP (this paper) in Fig. \ref{fig:php_time}A. We find that  the number of Fock-build calls   is reduced by at least a factor of 10 at each geometry (and sometimes more). This difference leads to a large savings in  time for performing the calculation; in in Fig. \ref{fig:php_time}, we plot the total wall time when running in  parallel with 48 cores.

\begin{figure*}[ht!]
    \centering
    \includegraphics[width=0.95\textwidth]{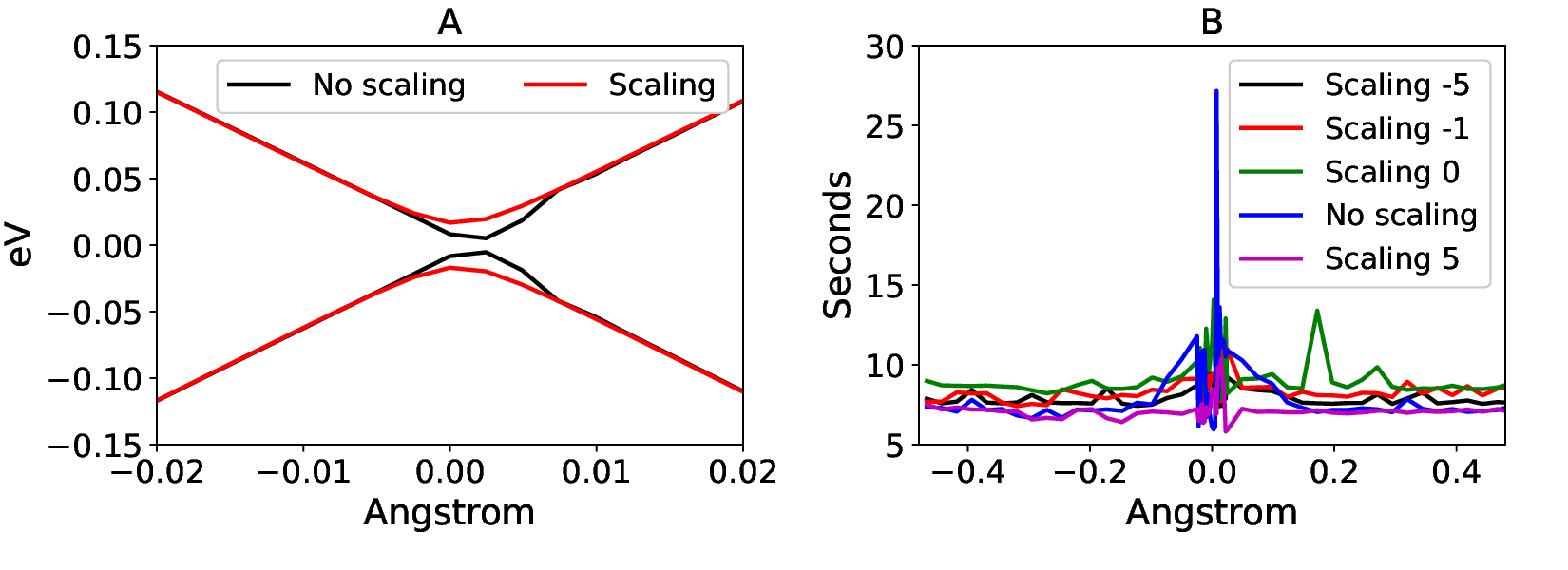}
    \caption{Potential energy surfaces (PES) and total wall time for different scaling factor of $\left(\tilde{\bm{F}}_\beta^2\right)_{N_d,N_d+1}$ when using DIIS-SQP for ph-ph system. (A) PES calculated with and without scaling, note that for a scaling factor of $-5,-1,0,5$, the PES are all identical to the red curve, while no scaling converges to the wrong states (see the lump on the black curve). (B) Total wall time given different scaling factors. The computational costs are comparable when a scaling factor is applied.}
    \label{fig:php_scale}
\end{figure*}

Let us next address the impact of 
scaling the coupling block between two active orbitals (i.e.  $\tilde{\bm{F}}_\beta^2$ from Eq. \ref{eq:scale_F_h}).  In Fig. \ref{fig:php_scale}, we plot results obtained when we performed  DIIS-SQP calculations with different scaling factors, namely $-5, -1, 0, 1{\rm (no\ scaling)},5$. In Fig. \ref{fig:php_scale}A shows that without a scaling factor, we cannot find the correct solution near the avoided crossing region (where $E_1$ and $E_2$ are close)  using the original error vector (Eq. \ref{eq:diis_err}).  However, with any non-unity scaling factor, simulations do onverge to the correct states, giving smooth potential energy surfaces (red curves in Fig. \ref{fig:php_scale}A). Interestingly, the failure of converging to the correct solution also causes more DIIS iteration steps, as shown in Fig. \ref{fig:php_scale}B. A detailed exploration suggests that the system spends too many steps trying to escape from the saddle point where $E_1 = E_2$. That being said, when a non-one scaling factor is applied, $E_1 = E_2$ is no longer a saddle point and the system can easily reach the desired minimum. Fig. \ref{fig:php_scale}B also suggests that the computational cost for different scaling factors is similar (as long as the scaling factor is not 1).

\subsection{am-ca (eDSC)}\label{sec:ac}
\begin{figure*}[ht!]
    \centering
    \includegraphics[width=0.95\textwidth]{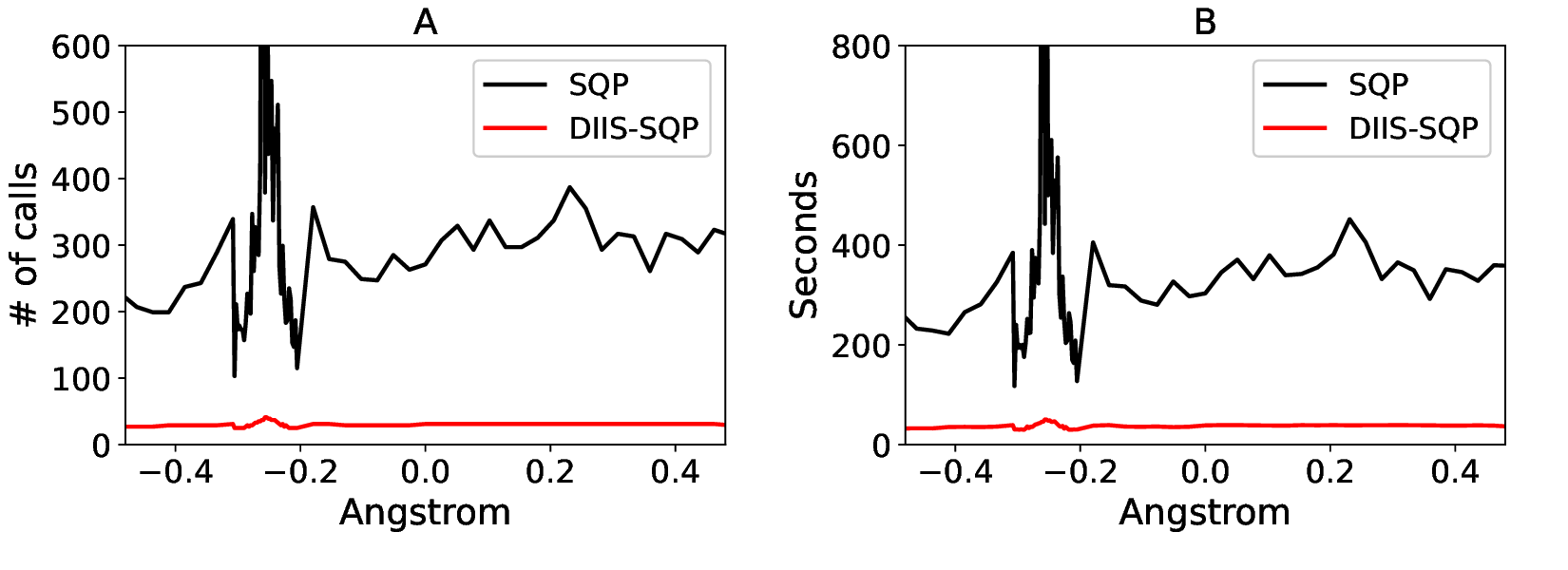}
    \caption{Computational cost of SQP and DIIS-SQP method for am-ca system. A scaling of -1 for $\left(\tilde{\bm{F}}_\beta^2\right)_{N_d+1,N_d+2}$ is applied for both methods. (A) Total number of calls to build the Fock matrices, i.e., $\bm{F}_\alpha^1,\bm{F}_\alpha^2,\bm{F}_\beta^1,\bm{F}_\beta^2$. These are the most time-consuming calls. (B) Total wall time for converging the gradient to less than $10^{-5}$ a.u. when running on 48 cores. DIIS-SQP reduces the computational cost by a factor of 8 to 20 compared with only using SQP.}
    \label{fig:ac_time}
\end{figure*}

\begin{figure*}[ht!]
    \centering
    \includegraphics[width=0.95\textwidth]{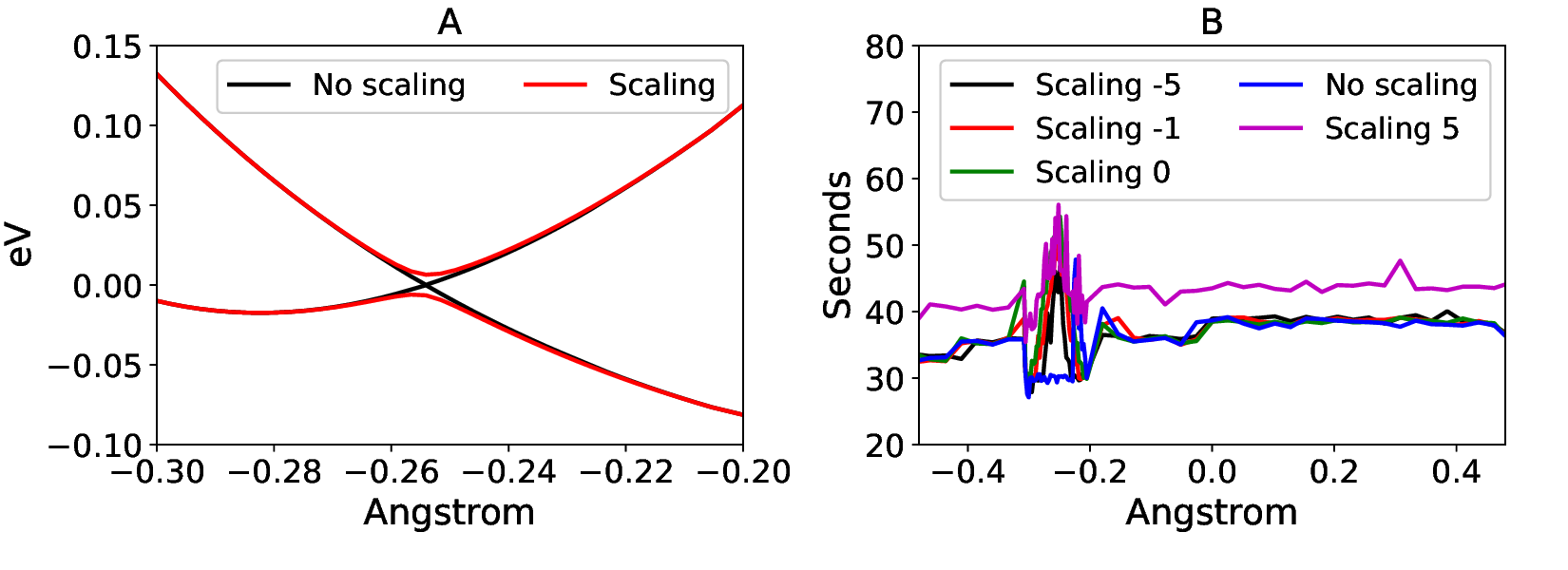}
    \caption{Potential energy surfaces (PES) and total wall time for different scaling factor of $\left(\tilde{\bm{F}}_\beta^2\right)_{N_d+1,N_d+2}$ when using DIIS-SQP for am-ca system. (A) PES calculated with and without scaling, note that for a scaling factor of $-5,-1,0,5$, the PES are all identical to the red curve, while no scaling converges to the wrong states (see the nearly crossing black curve in the avoided crossing region). (B) Total wall time given different scaling factors. The computational costs are comparable when a scaling factor is applied.}
    \label{fig:ac_scale}
\end{figure*}

Next, let us turn to the am-ca system, where we find similar conclusion. In Fig. \ref{fig:ac_time}A, we plot the number of  Fock-build calls for DIIS-SQP, which is reduced by at least a factor of 8 compared with SQP, and consequently, the total wall time is reduced by the same factor (see Fig. \ref{fig:ac_time}B). In terms of the scaling factors for $\left(\tilde{\bm{F}}_\beta^2\right)_{N_d+1,N_d+2}$ in Eq. \ref{eq:scale_F}, we again observe that non-unity scaling yields a smooth avoided crossing PES while a lack of scaling leads to an undesired PES (see Fig. \ref{fig:ac_scale}A). One small difference between the am-ca system and the ph-ph system is that the scaling does not lower the computational cost (see the blue curve Fig. \ref{fig:ac_scale}B). We note that, however, this finding is in fact artificial because, in practice,  the energy differences between two states near the avoided crossing region is small enough such that the un-scaled 
%gradient near the saddle point is less than our threshold ($10^{-5}$ a.u.) and hence the 
algorithm terminates (with a gradient threshold of $10^{-5}$ a.u.) before trying to escape from the saddle point. Another difference shown in Fig. \ref{fig:ac_scale}B is that using a scaling factor of 5 actually increases the computational cost by approximately 30\%-50\%. The reason for this increase is still unclear  -- perhaps a larger scaling factor requires convergence of  the gradient to a smaller threshold --but in any event,  a scaling factor of -1 is always helpful.

\section{Discussion}\label{sec:discussion}

\subsection{Cost compare to ROHF}
Our belief is that the current algorithm for eDSC/hDSC should be immediately capable of running meaningful calculations as a substitute for more accepted methods. To that end, in this section, we will demonstrate that DIIS-SQP can be only twice as expensive as a single-point ROHF calculation.  To show this,  we have applied ROHF to both the ph-ph and am-ca systems with the same reaction coordinates as in Figs. \ref{fig:php_time} and \ref{fig:ac_time}. For all simulations, the convergence criterion is that the errors in the total energy be less than $10^{-9}$ Hartree.  

\begin{figure*}[ht!]
    \centering
    \includegraphics[width=0.95\textwidth]{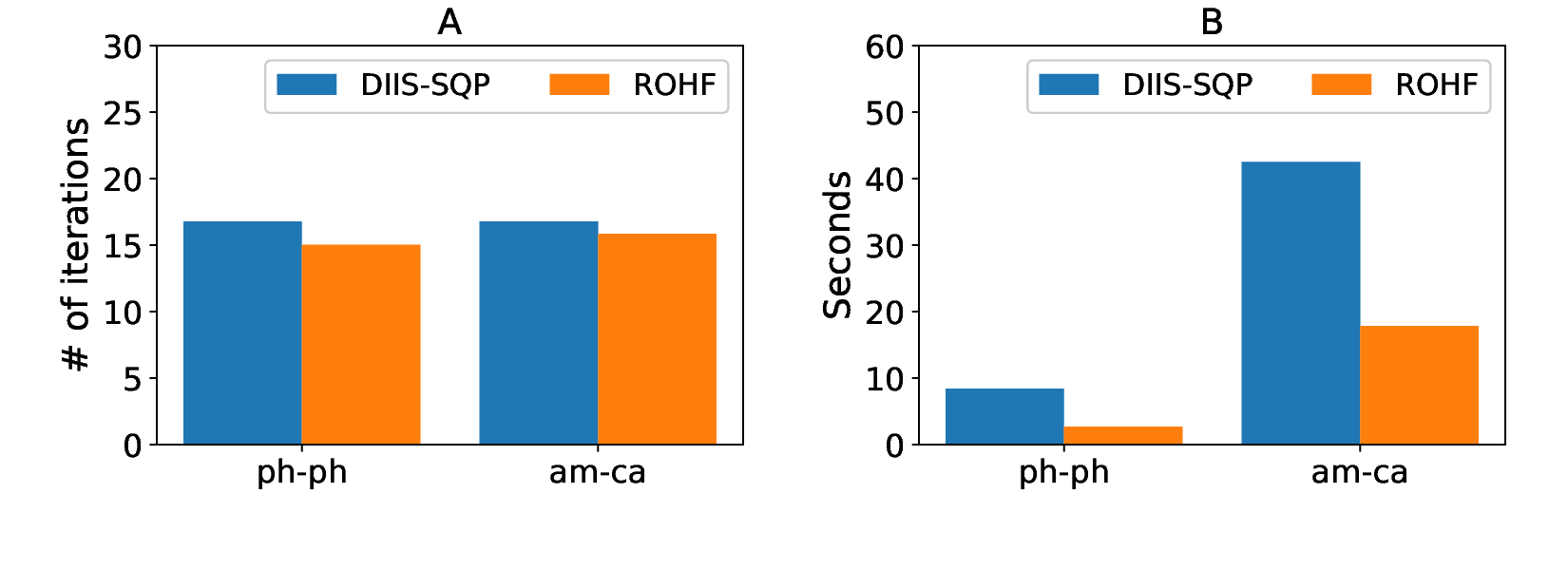}
    \caption{Comparison of computational cost between DIIS-SQP and ROHF. The convergence criterion is that errors in energy is less than $10^{-9}$ Hartree. For both DIIS-SQP and ROHF, the converged solution at previous geometry is read as the initial guess. (A) Averaged total number of SCF iterations. (B) Averaged total wall time to converge SCF iterations when running on 48 cores.}
    \label{fig:compare_rohf}
\end{figure*}

For both DIIS-SQP and ROHF, the converged solution at a previous geometry along the reaction path is read as the initial guess. The results are shown in Fig. \ref{fig:compare_rohf}. Clearly, Fig. \ref{fig:compare_rohf}A shows that DIIS-SQP requires almost the same number of SCF iterations as ROHF does regardless of the size of the system. Since two configurations are involved in DIIS-SQP and one needs to build the Fock matrices for both configurations, the total wall time is roughly twice the wall time for ROHF, as shown in \ref{fig:compare_rohf}B. Looking forward, we expect that we will be able to extend the current approach to treat systems having $N$ charge centers involved in a CT process, where $N$ configurations are required.  In that case, we also expect the total computational cost can be as little as $N$ times the cost of a single-point ROHF calculation (with the added benefit, of course, of generating $N$ electronic states rather than one), and these calculations will be of immediate use for semiclassical calculations exploring charge transport\cite{blumberger:2016:fragment_fssh_bo, prezhdo:2016:fragment_fssh_bo}.

\subsection{Cost of nSCF (SQP) steps}
\begin{figure*}[ht!]
    \centering
    \includegraphics[width=0.95\textwidth]{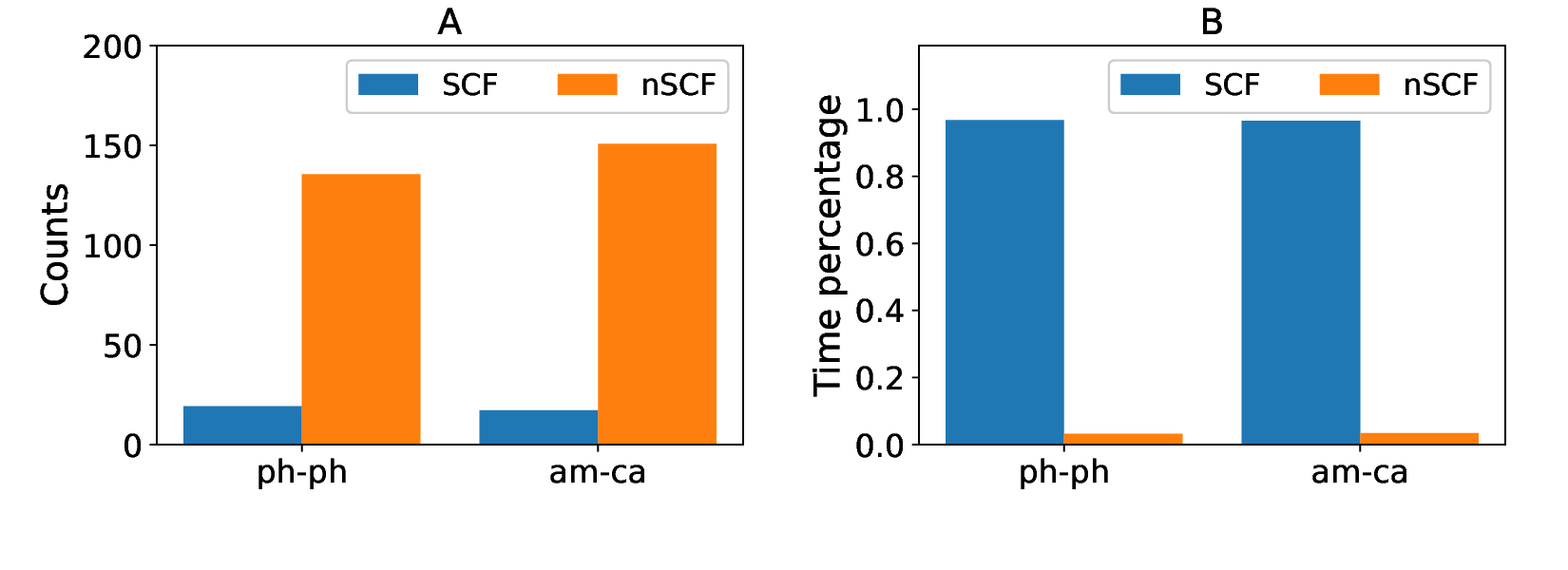}
    \caption{Cost of SCF (DIIS) steps and nSCF (SQP) steps with a scaling factor of -1. (A) Averaged total number of iterations for SCF and nSCF steps. (B) Averaged total wall time percentage for SCF and nSCF steps.}
    \label{fig:scf_nscf}
\end{figure*}

Finally, let us further investigate how and why the present algorithm is so much more efficient than the previous algorithm  by decomposing
the wall time between SCF and non-SCF time components.  Obviously, according to Figs. \ref{fig:php_time} and \ref{fig:ac_time}, 
building Fock matrices in the DIIS step (Step \ref{step:build_fock} in Sec. \ref{sec:algorithm}) dominates the total computational cost.
That being said, it is also true that the SQP steps for solving the nSCF problem do also come at a cost, which is important to recognize.
As shown in Fig. \ref{fig:scf_nscf}A, with our adaptive threshold (i.e., norm $(\bm{V}^n)/100$ in Step \ref{step:nscf}) applied to SQP for the nSCF problems, we find that an order of 10 to less than 100 nSCF steps are needed per SCF step. This finding implies that the most time consuming step in nSCF iterations, i.e., to calculate the exponential of a matrix, requires far less time consuming than building the Fock matrices (apparently less than 5\% of the total wall time, see Fig. \ref{fig:scf_nscf}B). It is worth noting that releasing the adaptive threshold to norm $(\bm{V}^n)/10$ reduces the required nSCF steps per SCF step, but the total number of SCF steps increases by approximately 40\% due to the inaccuracy of nSCF solutions, which would actually increase the total all time. Vice versa, if we were to 
strengthen the adaptive threshold to norm $(\bm{V}^n)/1000$, we find that will increase the number of nSCF steps per SCF step but not reduce the total number of SCF steps. For this reason, we believe the norm $(\bm{V}^n)/100$ appears to be a sweet spot as the adaptive convergence threshold for the nSCF steps.

% \subsection{An alternative DIIS error}
% Before concluding this paper, let us note that the expression in Eq. \ref{eq:diis_err} is not the unique way of defining the DIIS error vector. In fact, Eq. \ref{eq:dEdA} indicates that an alternative DIIS error vector can be defined as
% \begin{align}
%     \bm{V}_{\rm DIIS} =[\bm{C}^\top\bm{M}_0\bm{C},\bm{K}_0]+[\bm{C}^\top(\bm{M}_1-\lambda \bm{Q})\bm{C},\bm{K}_1]+[\bm{C}^\top(\bm{M}_2-\lambda \bm{Q})\bm{C},\bm{K}_2]\label{eq:diis_err_v2}
% \end{align}
% which is the gradient of $E_{\rm tot}$ w.r.t. the rotation generator $\bm{A}$ given the constraint in Eq. \ref{eq:constraint_q}. As a matter of curiosity, we have compared convergence behavior using Eq. \ref{eq:diis_err} and \ref{eq:diis_err_v2} and found no significant evidence that one DIIS error vector is better or superior than the other, as shown in Fig. \ref{fig:diis_v1v2}. As such, either Eq. \ref{eq:diis_err} or Eq. \ref{eq:diis_err_v2} can be used as the DIIS error vector.

% \begin{figure*}[ht!]
%     \centering
%     \includegraphics[width=0.95\textwidth]{diis_err_ana.eps}
%     \caption{Number of iterations for using different DIIS error vectors. Eq. \ref{eq:diis_err} is labeled as ``err\_v1'' an Eq. \ref{eq:diis_err_v2} is labeled as ``err\_v2''. (A) Averaged total number of iterations for SCF steps. (B) Averaged total number of iterations for nSCF steps.}
%     \label{fig:diis_v1v2}
% \end{figure*}

\section{Conclusion}\label{sec:conclusion}
In this paper, we have presented an efficient, production-ready algorithm for converging constrained cCASSDF(1,2) and cCASSCF(3,2) electronic structure calculations (so called eDSC and hDSC calculations). 
By separating the constrained minimization problem into an unconstrained SCF problem and an constrained nSCF problem, and by using DIIS to solve the unconstrained SCF problem, we have significantly reduced the number of Fock-build calls and hence minimized the total computational cost of eDSC/hDSC by a factor of 8 to 20 relative to Ref. \citenum{Qiu:2024:dsc}. With this new algorithm, we find that a eDSC/hDSC calculations is only twice as much more computationally expensive than a single-point ROHF calculation for the same system, making eDSC/hDSC a powerful for studying CT problems with non-adiabatic dynamics. We have also found that one can bypass a saddle point in eDSC/hDSC by simply scaling one matrix element in one component of the Fock matrix (Eq. \ref{eq:scale_F} and its equivalence for hDSC) by a factor of -1. Looking forward, the next step is clearly to generate gradients and derivative couplings, after which, propagating nonadiabatic dynamics should be readily possible. Future work will also consider replacing HF with DFT so as to treat molecules interacting with metal clusters, which represents a difficult but crucial interface problem for modern chemistry.\cite{Liu2023}

\section{Acknowledgement}
This work was supported by the U.S. Air Force Office of Scientific Research (USAFOSR) under Grant Nos. FA9550-23-1-0368 and FA9550-18-1-420.

\section{Supporting Information}
Geometries of ph-ph and am-ca; convergence performance comparison: finite versus infinite temperature

\bibliography{cite.bib}

%\newpage
%\section{For Table of Contents Use Only}
%\begin{figure*}[ht!]
%    \centering
%    \includegraphics[width=0.5\textwidth]{TOC.png}
%    \caption{TOC}
%    \label{fig:toc}
%\end{figure*}

\end{document}